\documentclass[a4paper,twocolumn]{esapub} 

\usepackage{epsfig}

\title{Explosive Nucleosynthesis}
\author[1,2]{M. Hernanz}
\affil[1]{Institut d'Estudis Espacials de Catalunya, IEEC,\cr
Edifici Nexus, C/Gran Capit\`a, 2-4, 08034 Barcelona, Spain}
\affil[2]{Instituto de Ciencias del Espacio, CSIC}

\newcommand{\gap}{\mathrel{ \rlap{\raise.5ex\hbox{$>$}}
                      {\lower.5ex\hbox{$\sim$}}  } }
\newcommand{\nil}{$^{56}$Ni}
\newcommand{\nim}{$^{57}$Ni}
\newcommand{\nih}{$^{60}$Ni}
\newcommand{\col}{$^{56}$Co}
\newcommand{\com}{$^{57}$Co}
\newcommand{\coh}{$^{60}$Co}
\newcommand{\fel}{$^{56}$Fe}
\newcommand{\fem}{$^{57}$Fe}
\newcommand{\feh}{$^{60}$Fe}
\newcommand{\tit}{$^{44}$Ti}
\newcommand{\sca}{$^{44}$Sc}
\newcommand{\caa}{$^{44}$Ca}
\newcommand{\all}{$^{26}$Al}
\newcommand{\mgg}{$^{26}$Mg}
\newcommand{\naa}{$^{22}$Na}
\newcommand{\neo}{$^{22}$Ne}  
\newcommand{\bee}{$^{7}$Be}
\newcommand{\lit}{$^{7}$Li}

\begin{document}

\maketitle


\begin{abstract}
Many radioactive nuclei relevant for gamma-ray astrophysics are synthesized
during explosive events, such as classical novae and supernovae. A review of
recent results of explosive nucleosynthesis in these scenarios will be
presented, with a special emphasis on the ensuing gamma-ray emission from
individual nova and supernova explosions. The influence of the dynamic
properties of the ejecta on the gamma-ray emission features, as well as the
still remaining uncertainties in nova and supernova modelling will also be
reviewed.                  
\end{abstract}

\section{Introduction}

In this paper I will review the sites of explosive nucleosynthesis relevant 
for gamma-ray astronomy. Two main types of explosion are responsible for the 
emission of gamma-rays in the Galaxy: supernovae, both thermonuclear and 
gravitational (core-collapse) and classical novae. 

Thermonuclear supernovae, or supernovae of type Ia, are exploding white dwarfs 
in close binary systems, which do not leave a remnant after the explosion. 
Core collapse supernovae (supernovae of type II, Ib/c) are exploding massive 
stars (M$\gap 10 {\rm M}_\odot$), which leave as remnant either a black hole 
or a neutron star. Typical velocities of supernovae ejecta are some $10^4$ 
km~s$^{-1}$, energies involved $10^{51}$ erg and ejected masses some  
M$_\odot$.

Classical novae are the result of the explosion of the 
external H-rich accreted shells of a white dwarf in a binary system. These 
explosions are recurrent phenomena (contrary to supernovae), since an 
explosion is expected every time the critical accreted mass on top of the 
white dwarf is reached. Typical velocities of novae ejecta are between 
some $10^2$ and $10^3$ km~s$^{-1}$, energies involved $10^{45}$ erg and 
ejected masses between $10^{-3}$ and $10^{-5}$ M$_\odot$.

The radioactive isotopes synthesized during explosive nucleosynthesis, 
either in novae or in supernovae, are summarized in table 1. Three types of 
decay chains can occur: electron captures (\nil \  $\rightarrow$ \ \col, 
\nim \  $\rightarrow$ \ \com \  $\rightarrow$ \ \fem,  
\tit \  $\rightarrow$ \ \sca\ and \bee \  $\rightarrow$ \ \lit), 
$\beta^+$ decays (\col \  $\rightarrow$ \ \fel, \sca \  $\rightarrow$ \ \caa, 
\all \  $\rightarrow$ \ \mgg\ and \naa \  $\rightarrow$ \ \neo) and 
$\beta^-$ decays (\feh \  $\rightarrow$ \ \coh \  $\rightarrow$ \ \nih). 
The first six isotopes in the table (\nil, \col, \nim, \tit, \all\ and \feh) 
are produced in supernova explosions (although not exclusively, at least in 
the case of \all), whereas the last two are synthesized in classical novae. 
In sections 2 and 3 below, I will discuss how the synthesis of these 
radioactive nuclei proceeds in these scenarios. In the case of core-collapse 
supernovae, it is important to distinguish the nucleosynthesis during the 
pre-explosive stage of the massive star evolution from that in the explosive 
phases; it is crucial as well to know which part of the star will finally be 
ejected, since this quantity will determine the final enrichment of the Galaxy 
in radioactive (and other) elements. I won't discuss the synthesis of 
radioactive isotopes in other (non-explosive) sites, like the AGB stars (see 
the contribution by Mowlavi, these proceedings) or the Wolf-Rayet stars 
(which can eject radioactive nuclei by strong stellar winds, during their 
hydrostatic evolution; see Arnould and Meynet 1997 and Meynet and Arnould, 
1999 for recent reviews).

\section{Synthesis of radioactive isotopes in supernova explosions}

Two types of isotopes can be distinguished, depending on their lifetime 
(see Diehl and Timmes, 1998, for a recent review). 
Short-lived isotopes, such as \nil, \nim\ (and their daughters \col\ and 
\com), \tit\ and \coh, have lifetimes short enough (see table 1) to make 
them detectable in individual objects. \nil\ and \nim\ are 
produced in all types of supernovae; \tit\ is mainly produced in 
core-collapse supernovae, but it can also be synthesized in thermonuclear 
supernovae of the sub-Chandrasekhar type (provided that they exist; see 
discussion of SNeIa types below). \coh\ is produced directly and from 
\feh\ decay, with \feh\ belonging to the long-lived isotopes group.

Long-lived radioactive isotopes, such as \all\ and \feh, 
have lifetimes long enough to make them undetectable in individual 
sources, because the nuclei can be quite far away from their source 
and mixed with those coming from other explosions (since the lifetime is 
longer than the typical period between two succesive explosions in the 
Galaxy). For these isotopes, only the accumulated emission in the Galaxy 
can be observed and used as diagnostic of models and of the Galactic 
distribution of the sources. The same classification scheme applies to 
isotopes synthesized in novae; in this case, \bee\ belongs to the 
short-lived group, whereas \naa\ belongs to both of them (see section 3 
below). 

A very recent and interesting compilation of papers about astronomy with 
radioactivities can be found in Diehl and Hartmann (1999).

\subsection{Observational clues}

Gamma-ray astronomy provides an unique opportunity to detect radioactive 
isotopes in individual objects, giving a proof of ongoing nucleosynthesis 
in them. In the case of supernovae, another tool for the determination of 
the amount of radioactive nuclei in the ejecta is the bolometric light 
curve (UVOIR, from {\sl ultraviolet, optical} and {\sl infrared}). In 
addition to the well known fact that \col\ (daughter of \nil) powers the 
early evolution of the
light curve (\nil\ mass can be determined from luminosity at maximum), \com\ 
is responsible for powering the light curve from around day 1000 after 
maximum. Later on \tit\ will provide a floor to the bolometric light curve and 
\naa\ and \coh\ could also play a minor role, depending on the supernova type 
and the specific yields of these radioactivities (see, e.g., Woosley, 
Pinto and Hartmann, 1989, Timmes et al. 1996, Diehl and Timmes 1998).

These two types of observational approaches to the radioactive content in the 
ejecta had been possible for only one object so far: the supernova 1987A, 
which exploded 13 years ago in the LMC, only 55 kpc away from us. This was a 
type II supernova, which is not the most favorable case to look for 
gamma-ray emission, since its is much more opaque and has a smaller content 
of the most relevant radioactivites than SNIa (see below); however, its 
very short 
distance allowed for detection of its gamma-ray emission and also for a 
follow-up until very late times of its light curve (through photometry in the 
UVBRIJHK bands).

Gamma-ray lines from \col\ decay, at 847 and 1238 keV, were detected 
in SN 1987A with the GRS instrument of the SMM satellite (Matz et al. 1988) 
and confirmed by 
several ballon-borne instruments (i.e., 
Teegarden et al. 1989, Mahoney et al. 1988, Sandie et al. 1988, Cook et al. 
1988, Rester et al. 1989). One surprising fact was the appearence of these 
lines only 200 days after the explosion (Matz et al., 1988), much earlier 
than expected. This has 
been interpreted as a sign of some early extra mixing of \col\, in order 
to transport this isotope into regions of low gamma-ray optical depth (see, 
e.g., Pinto and Woosley, 1988, Leising 1988, Bussard, Burrows and The, 
1989, Leising and Share, 1990). The line profiles observed with GRIS 
(Teegarden et al. 1989, Tueller et al. 1990) have also put constraints on 
theoretical models of supernova explosions, since sphericity and homogeneity 
of the ejecta were incompatible with the observed fluxes and widths of the 
\col\ lines. 

Another crucial gamma-ray observation of short-lived isotopes in SN 1987A 
was the detection of gamma-ray radiation from \com\ decay (between 50 and 
136 keV), with OSSE on the CGRO (Kurfess et al. 1992). The deduced \com\ 
content (for models with low gamma-ray optical depth, see Kurfess et al. 
1992) was such that the original ratio \nim/\nil\ produced in the explosion 
should be about 1.5 times the solar \fem/\fel\ ratio . Observations up to now 
show the change of slope related to the sequence of \col-\com\ decays 
(see figure 3 in Timmes et al. 1996). Future UVOIR observations would possibly 
be able to show the light curve powering from \tit. The SPI instrument onboard 
INTEGRAL has some possibilities to detect the gamma-ray emission from this 
\tit, which would provide a unique proof of the nucleosynthesis in core 
collapse supernovae and an important link between the UVOIR and the 
gamma-ray observations. Coming back to \com\, it was first thought that 
the SN 1987A 
\com\ content deduced from OSSE observations was not enough to power the
available bolometric light curve, since 5 times solar \fem/\fel\ ratio was 
required (Suntzeff et al. 1992) and alternative mechanisms to power the 
bolometric light curve were suggested (Clayton et al. 1992). However, more 
recent observations seem to 
require a smaller amount of \com-decay to power the light curve, in agreement 
with the \fem/\fel\ ratio deduced from OSSE observations (see figure 9 in
Diehl and Timmes 1998).

The excitement induced by the above mentioned gamma-ray observations 
(together with many other observations at other wavelength ranges) has led the 
theorists to suggest different possibilities for mixing both during the 
explosion and the ejection phases (to quote only a few of the early works, 
see e.g., Arnett, Fryxell 
and M\"uller 1989, Benz and Thielemann 1990, Fryxell, Arnett and M\"uller 
1991, Herant and Benz 1992, 
and also the general reviews of SN 1987A from Arnett et al. 1989 and 
McCray 1993 and references therein). 

Another detection (tentative) of gamma-ray emission  from a supernova, 
with COMPTEL on CGRO, was that of SN 1991T (which was an overluminous SNIa), 
in NGC 4527 at around 17 Mpc distance. In that case, a marginal detection 
of the 847 keV line 
was reported (Morris et al. 1995, 1997), leading to a prediction of a 
quite large \nil\ mass, implying that all the white dwarf mass should have 
been incinerated to \nil\ (in contradiction with current theoretical models). 
More recently, upper limits to the fluxes of the 
847 and 1238 keV lines from \com-decay in the type Ia supernova 
1998bu, in NGC 3368 at around 8 Mpc distance, have been deduced from 
COMPTEL observations (Georgii et al. 2000). Although no detection has 
been obtained, these limits are restrictive enough to constrain some of 
the available models of SNeIa nucleosynthesis (like a sub-Chandrasekhar mass 
model from Nomoto et al. 1997).

There is another important observation of gamma-ray lines related to short and 
medium-lived radioactivities in supernovae: the discovery of \tit\ emission 
at 1157 keV in the Cas A 
supernova remnant (Iyudin et al. 1994; see reviews from Diehl and Timmes 
1998 and Kn\"odlseder, these proceedings). Again the observations in 
gamma-rays are in 
some way puzzling, because the amount of \tit\ deduced from observations 
implies a \nil\ content (according to theoretical models of supernovae
nucleosynthesis) which should have originated a very bright supernova, in 
contrast with the absence of historical records (Timmes et al. 1996). 
Observations in gamma-rays push forward the 
theoretical models, in order to balance all the available possibilities 
and to consider new ones.

A different kind of information is obtained from the observations of 
long-lived radioactivities (\all\ and \feh). In this case, what is seen 
is not the ongoing nucleosynthesis in a particular object, but the integrated 
nucleosynthesis in the Galaxy. Up to now, this has been possible for 
the 1809 keV \all\ emission. The \all\ map obtained with the COMPTEL 
instrument onboard the Compton Gamma-Ray Observatory CGRO (Diehl et al. 1995, 
1997, Oberlack et al. 1996, Kn\"odlseder 1997, 1999,  
Pl\"uschke et al. these proceedings) has posed interesting questions 
about the origin of the galactic \all\ (see, e.g., review 
from Prantzos and Diehl, 1996). It 
provides a direct and unique insight on the integrated nucleosynthesis during 
the last $10^6$ years. Some regions of enhanced emission have been discovered 
(Cygnus, Carina, Vela), indicating the presumable link between \all\ emission 
and massive star formation, as well as the relationship with spiral structure 
of the Galaxy (Diehl et al. 1996, Kn\"odleseder et al. 1996a,b, 
Diehl et al. 1999, Kn\"odlseder, these proceedings). 
For \feh, a similar map should be observed by INTEGRAL, because the sources 
of this isotope are the same as those of \all, being the yields smaller by 
some factor (Timmes et al. 1995, Diehl et al., 1997). 

It is worth mentioning that the integrated nucleosynthesis of \tit\ and 
\naa\ may also be seen, if instruments are sensitive enough. The 
future \tit\ and \naa\ maps will provide a precious information about their 
sources, i.e., supernovae (mainly core collapse ones) and novae, respectively.

\subsection{Thermonuclear supernovae (SNeIa)}

The defining characteristic of SNeIa is the lack of hydrogen in their spectra, 
as well as the presence of a P Cygni feature related to SiII, 
$\lambda6335$, at maximum light (Wheeler 
and Harkness 1990); in general, intermediate-mass elements (O, Mg, Si, S, Ar, 
Ca) appear in the spectrum near maximum light with high velocities 
(8000-30000 km~s$^{-1}$). SNeIa are quite homogeneous from 
the observational point 
of view (i.e, $\sim$90\% of all SNeIa have similar spectra, light curves and 
peak absolute magnitudes), although some differences exist (i.e., subluminous 
explosions, like SN1991bg and SN1992K, and overluminous ones, like SN1991T).
SNeIa appear in both elliptical and spiral 
galaxies and, therefore, their progenitors should be long-lived. These facts 
all together suggest that the thermonuclear disruption of 
mass-accreting carbon-oxygen (CO) white dwarfs is responsible for these 
explosions. 
Already in the sixties, Hoyle and Fowler (1960) suggested that thermonuclear 
burning in an electron-degenerate stellar core might be responsible for 
type I supernova (there was no subclassification at the epoch) explosions, 
with  the explosion energy coming from the thermonucler burning of CO into 
higher mass elements (see also, e.g., the pioneering works by Arnett 1969, 
Hansen and Wheeler 1969). It was also suggested at the epoch that the early 
supernova luminosity might 
have its origin on the radioactive decay of \nil\ (Colgate and McKee 1969), 
which was already known to be a product of supernova nucleosynthesis, and 
that gamma-ray lines should be emitted from those explosions (Clayton, 
Colgate and Fishman, 1969). But the particular {\it scenario} 
where the explosion occurs (see, e.g., Livio 1999) and the  
{\it physics of the flame} itself (see, e.g., Hillebrandt and 
Niemeyer 2000) are far from being understood.

Two types of progenitors have been suggested so far, concerning the mass of 
the 
exploding CO white dwarf: Chandrasekhar and sub-Chandrasekhar mass models. In 
the Chandrasekhar mass models, a CO white dwarf explodes when reaching that 
mass, with central carbon ignition propagating outwards being 
responsible for the explosion. The main problems related to this model are 
the uncertainties concerning burning propagation (deflagration,  
detonation, delayed detonation, see below), but also the scenario is 
unclear. Either a double degenerate (merging of two CO white dwarfs) or a 
single degenerate scenario is possible. In all cases, the growth to the 
Chandrasekhar mass is problematic, because both mass loss (through nova 
episodes, for instance)  and accretion induced collapse (if 
the initial mass is high enough and/or the white dwarf is made of oxygen and 
neon) should be avoided (see, e.g., Canal, Isern and Labay 1990, 1992, 
Canal et al. 1990, Isern, Canal and Labay 1991, Nomoto and Kondo 1991,
Bravo and Garc\'{\i}a-Senz 1999). 
In addition, for the double degenerate scenario there is a problem 
of statistics: there are not enough double white dwarf systems with 
sufficiently short period and total mass in excess of the Chandrasekhar mass 
able to explode in less than the Hubble time and to explain the galactic 
SNeIa rate. In fact there wasn't any observed system fulfilling these 
conditions 
until the very recent discovery of KPD 1939+2752 (Maxted, Marsh and North, 
2000), which is the first SNIa progenitor candidate observed. 

In the sub-Chandrasekhar mass models, a CO white dwarf of low-mass 
(0.6-0.8 M$_\odot$) accretes helium 
($\Delta\rm M_{\rm He} \sim 0.1-0.2$ M$_\odot$), 
reaching a final mass smaller than the Chandrasekhar mass. Provided the 
accretion rate is moderate (around $10^{-8}$M$_\odot~{\rm yr}^{-1}$), there 
is He ignition on the top of the CO core. This ignition causes 
an outward propagating He-detonation wave (basically transforming He into Ni 
at high velocity) and an inward propagating pressure 
wave. The last one finally provoques a carbon ignition (central or 
off-center), which leads to an outward carbon-detonation incinerating all the 
white dwarf, and synthesizing intermediate-mass elements, in addition to Ni 
(see, e.g., Livne 1990, Livne and Glasner 1991, Woosley and Weaver 1994). 
Therefore, in this model (called ``indirect double detonation'', IDD, or 
``edge lit detonation'', ELD) there is an outer layer of high-velocity Ni 
and He above the 
intermediate-mass elements, which does not exist in the Chandrasekhar-mass 
models. Sub-Chandrasekar mass models are not considered as 
good SNeIa progenitors nowadays, because of both observational and theoretical 
problems; observational: the high velocity Ni above intermediate mass 
elements is not seen in the spectra; theoretical: the He-driven carbon 
detonation is very model dependent (see for instance the 3D models from 
Garc\'{\i}a-Senz, Bravo and Woosley 1999). But it is still a possibility that 
sub-Chandrasekar mass models explain some subluminous SNeIa, like SN1991bg 
(see, e.g., Ruiz-Lapuente, Canal and Burkert 1997). 

In summary, the bulk of normal SNeIa are assumed to be exploding 
Chandrasekhar-mass CO white dwarfs, but there is still room for the 
sub-Chandrasekhar mass models to explain some peculiar objects. Therefore, 
whether SNeIa come from single or double-degenerate scenarios and whether 
they come from carbon or helium plus carbon ignition are not closed issues 
(see, e.g., the recent paper from Branch 2000). 

The main problems still remaining on the modeling of SNeIa affect the ignition 
process and the flame propagation. Different possibilities exist: deflagration
(subsonic flame speed), detonation (supersonic) and a combination of both 
(delayed detonation). A detonation with densities larger than $\sim 10^7$ 
g~cm$^{-3}$ is not a viable mechanism, since all the star would be 
incinerated to Ni, without synthesis of intermediate-mass elements. On the 
contrary, if the density is lower, intermediate-mass elements are 
synthesized, in agreement with the observations. Concerning deflagrations, 
they produce nucleosynthesis at velocities in general agreement with the 
observed spectra, but some neutronized isotopes (such as $^{54}$Fe, $^{54}$Cr 
and $^{58}$Ni) are overproduced in amounts incompatible with the chemical 
evolution of the Galaxy. To overcome this problem, delayed detonations were 
suggested (Khokhlov 1991a). There are two 
situations in which a deflagration to detonation transition (DDT) could occur 
in supernovae
(see, e.g., Khokhlov, Oran and Wheeler, 1997): DDT could occur directly 
or as a result of a previous expansion. For instance, in 
the pulsation delayed detonation, a first slow deflagration is quenched 
because of the expansion of the white dwarf, 
which subsequently pulses and recontracts, causing a detonation upon 
recollapse (Khokhlov 1991b). The propagation of the 
detonation wave through the pre-expanded star produces the required 
intermediate mass elements in the outer layers at densities lower than 
$\sim 10^7$g~cm$^{-3}$ (which are not synthesized in detonations at larger 
densities). In these models, the problem of overproduction of 
highly-neutronized nuclei is alleviated but not solved (Khokhlov 1991a, b). 
Models of delayed detonations in 2D, both of the first deflagration phase and 
of the subsequent detonation phase, have been performed by Arnett and Livne 
(1994a, b); they show that the first slow deflagration is insufficient to 
unbind the star, that a pulsation of large amplitude is generated and that 
reignition occurs after the first contraction phase.

In summary, there is a 
general consensus about the fact that, in order to explain spectroscopic 
observations, burning should proceed subsonically (deflagration) in the inner 
core (where densities are large, i.e., $\rho > 10^8$ g~cm$^{-3}$), whereas 
burning becomes supersonic (detonation) in the outer lower density zones 
(see examples of models in Bravo et al. 1993, H\"oflich and Khokhlov 1996, 
Bravo et al. 1996, Woosley 1997).
But the way in which the deflagration to detonation transition (DDT) occurs 
is not yet clear, (see, e.g., discussions in recent papers by Niemeyer and 
Woosley 1997, Niemeyer 1999, Lisewski, Hillebrandt and Woosley 2000, and 
in the review by Hillebrandt and Niemeyer 2000). There is also ample debate 
about the way in which the initial burning occurs: flame instabilities, 
flame-turbulence interactions (see review about turbulence and thermonuclear 
burning by Hillebrandt and Niemeyer 1997, and references therein).

All 1D models (which were the unique ones available up to 
the nineties and still are the only ones to include {\it complete 
nucleosynthesis}) 
rely on prescriptions based on some parametrization of the flame speed and,  
in the case of delayed detonations, of the deflagration-detonation transition 
-DDT- densities. Different groups work in models of thermonuclear SNIa and 
their nucleosynthesis, including the radioactivities. It is out of the 
scope of 
this paper to mention even a small fraction of them, but a small sample 
can be useful to show the main results and the main caveats still remaining 
(see the recent books {\it Thermonuclear Supernovae}, edited by Ruiz-Lapuente, 
Canal and Isern, 1997, and {\it Type Ia Supernovae: Theory and Cosmology}, 
edited by Niemeyer and Truran, 2000).
Nomoto and coworkers have computed detailed nucleosynthesis in carbon 
deflagration supernovae (Nomoto, Thielemann and Yokoi, 1984, Thielemann, 
Nomoto and Yokoi, 1986), and also in other types of explosive carbon burning 
(such as delayed detonations, with parametrized ignition densities and 
deflagration-detonation transition -DDT- densities, see Iwamoto et al. 1999). 
The yields of radioactive isotopes are mainly affected by the DDT density 
(i.e., synthesized mass of \nil\ ranges from 0.55 to 0.77 M$_\odot$, and 
\nim\ from 
9.6x$10^{-3}$ to 1.98x$10^{-2}$ M$_\odot$, in Iwamoto et al.'s models). These 
yields are larger than those from core collapse supernovae (see below) and 
distributed in less opaque zones, since there isn't much mass above them.
This makes type Ia supernovae better targets for INTEGRAL than SNeII (but 
see section 2.1 for observational results).
In the context of gamma-ray astronomy, it is important to stress that 
sub-Chandrasekhar mass models synthesize larger amounts of \tit\ than 
Chandrasekhar mass ones (see, e.g., Woosley and Weaver, 1994). 

Gamma-ray spectra of SNeIa for the different models provide important 
signatures of the explosion mechanism, although unfortunately there isn't 
much observational data to compare with (see previous section). Prospects for 
SNeIa explosion mechanism identification with gamma-rays have been analyzed 
recently by G\'omez-Gomar et al. (1998a), with a special emphasis 
on detectability with the instruments that will be onboard INTEGRAL (see also 
Burrows and The 1990, H\"oflich, Khokhlov and M\"uller 1994, Kumagai and 
Nomoto 1997, H\"oflich, Wheeler and Khokhlov 1998). 

Lines from \nil-decay (158, 750, 812 keV) are prominent during the first days 
after the explosion, but they disappear very fast, because of the short 
\nil-lifetime. Lines from \col\ (847, 1238 keV) and \com\ (122, 136 keV) 
appear later and have longer durations. The most intense lines are those 
at 847, 1238, 812 and 158 keV, in addition to the annihilation line at 511 
keV. The strongest line is always the 847 keV one (detectable up to 11-16 Mpc 
with SPI on INTEGRAL), whereas the 158 keV line (from \nil-decay) is the 
most interesting to discriminate between models. The 158 keV line is narrower 
and, therefore, detectable at longer distances with SPI, than another 
\nil-line (at 812 keV), despite being fainter. It is almost undetectable 
in pure deflagration models, whereas it is even stronger than the 1238 line 
in detonation models. Another interesting 
signature of the models is the ratio between the 847 keV and the 158 keV line 
fluxes (200 days after maximum and at maximum, respectively), because it 
provides information about the ratio between total \nil\ in the ejecta and 
\nil\ in the external layers: the late emission at 847 keV comes from 
\col-decay (coming from \nil-decay), while only the \nil\ present in the 
outermost shells is responsible for the 158 keV line flux (see G\'omez-Gomar 
et al. 1998a for details). Finally, line profiles will also 
provide important information allowing for discrimination between the models, 
for explosions at distances short enough (see again G\'omez-Gomar et al. 
1998a).

\subsection{Core-collapse supernovae}

All supernova types except type Ia's (i.e., type II, Ib/c) are explained by 
the explosion of massive stars. Stars with initial masses 
(M$ \gap 10$ M$_\odot$) 
don't end their lives as white dwarfs. Succesive phases of thermonuclear 
burning (C, Ne, O, Si) give as a result a star with an ``onion-skin'' 
structure, where a central iron core is surrounded by shells made of elements 
of progressively lower atomic mass. The  
chemical composition along the star is the following (see, e.g., figure 10.8, 
corresponding to a 25 M$_\odot$ star, in Arnett 1996): Fe core, 
{\it Si-burning} zone (made mainly of elements from Si to Ni, without O), 
{\it O-burning} zone (O, Si-Ca), {\it Ne-burning zone} (Ne, Mg and O, no C), 
{\it C-burning} zone (C and O, Ne and Mg), radiative He-burning zone 
(He, C and O), convective He zone, inert part of old He core (interior to 
H-burning shell), material above the H-burning shell (plus some inert 
zones associated with the Si, O, Ne and C-burning zones and just outside 
them). Once the Fe-core reaches the Chandrasekhar mass, it becomes unstable 
and collapses to form a neutron star. The gravitational energy released 
($\sim 10^{54}$ erg) during core collapse is 
responsible for the ensuing supernova explosion, but it is not yet 
completely understood how the conversion of this potential energy into 
kinetic energy proceeds (only 0.1\% of the available potential energy is 
needed). 
 
Baade and Zwicky (1934) were the first to suggest that the gravitational 
energy released during the formation of a neutron star could produce a 
supernova explosion. Colgate and White (1966) built a supernova model, 
considering 
that the transfer of energy takes place by the emission and deposition of 
neutrinos; Wilson (1971) showed that the electron capture neutrino burst 
was not strong enough to eject material. The Weinberg-Salam model of 
electroweak interactions opened new possibilities of neutrino interactions 
with matter (neutral currents). 
In 1974, Freedman noticed the importance of neutral currents in 
the physics of core collapse supernovae; as a result of the increased cross 
section of core material to neutrinos, these particles are trapped during the 
collapse. It was shown by Bethe et al. in 1979 that one of the consequences of 
neutrino trapping is that the entropy of the core changes little during 
collapse (it remains low), leaving the collapse continue up to nuclear 
densities. Further compression is prevented by the repulsive component of 
the strong interaction (stiffness of nuclear matter), leading to the core 
bounce. A shock wave is generated at its boundary and propagates outwards. 
But it has been shown that the energy of this shock is mainly invested in 
the photodisintegration of heavy nuclei and in neutrino losses; therefore, 
the shock stalls and the explosion via the so called ``prompt mechanism'' is 
unsuccessful. In the ``delayed mechanism'', there is a revival of the stalled 
shock because of neutrino heating behind the shock (Bethe and Wilson 1985). 
However, the explosion energy does not reach easily the 
necessary $10^{51}$ erg. Further works introduced the effect of convective 
instabilities, caused by a negative entropy gradient, in order to deliver 
energy to the shock (see, e.g., Bethe, 
1990, Herant, Benz and Colgate 1992, Herant et al. 1994, Bethe 1995, Janka 
and M\"uller, 1995, Burrows, Hayes and Fryxell, 1995, to quote only a few of 
the papers dealing with this topic). 
Convection aids the explosion because it increases the efficiency at which 
neutrino energy is deposited (material that rises cools and converts energy 
from neutrino deposition into kinetic energy, instead of re-radiating it as 
neutrinos) and also reduces the energy required to launch the explosion (by 
reducing the pressure at the accretion shock) (see recent reviews by Fryer, 
2000, Burrows 2000, and references therein). The handling of this process is 
very model dependent: treatement of neutrino transport, multidimensional 
aspects. Also the structure of the stellar core before collapse (i.e.,  
the presupernova model) are important for the final outcome of the explosion.

Fortunately, nucleosynthesis during core collapse supernova explosions can 
be computed without a complete knowledge of the explosion mechanism itself. 
As in the case of thermonuclear supernovae, all the details of the physics 
involved in the explosion are not required to have an approximate, but 
quite good 
when compared with the observations, idea of which are the main 
nucleosynthetic yields of core collapse supernovae. Two steps are needed to 
compute SNII (and Ib/c) yields: nucleosynthesis during the massive star 
evolution (i.e., pre-supernova phase) and explosive burning when a shock wave 
crosses the mantle surrounding the collapsing core. 

There are different ways to simulate the explosion artificially. One is by 
means of a ``thermal bomb'', i.e., injecting thermal energy inside the Fe 
core, in a way such that the ejecta attains the desired kinetic energy, 
$\sim 10^{51}$erg (see, e.g., Thielemann, Nomoto and Hashimoto, 1996). Another 
alternative is the injection of momentum, through a piston, inward-moving 
during the infall previous to the explosion, and outward-moving during the 
explosion, with a velocity such that the desired kinetic energy of the ejecta 
is obtained (see, e.g., Woosley and Weaver, 1995). The {\it mass cut} 
between the collapsing core and the ejecta determines the amount of mass 
ejected (and that of \nil\ and other radioactive isotopes, in particular). In 
the ``thermal bomb'' method, they adjust it taking into account the 
relationship between supernova progenitor masses and \nil\ masses ejected 
deduced from some 
observations. In the piston approach, the mass cut is obtained from the 
choice of piston position and energy; a mass cut located outside the 
piston is often obtained (for a discussion of the differences between both 
models, including an analysis of the influence of the nuclear reaction rates, 
see Hoffman et al., 1999). In summary, both groups have performed calculations 
of detailed nucleosynthesis by inducing the core-collapse supernova explosion 
on massive stars (previously evolved following all the nucleosynthesis 
phases). Other groups have performed studies of massive star evolution, but 
there is no room in this short review to mention all of them. 

The masses studied by Thielemann et al. (1996) range between 13 and 25 
M$_\odot$, with initial metallicities, Z, equal to solar (see Nakamura et al. 
1999 for the effect of low Z). Woosley and Weaver (1995) studied the range 
11-40 M$_\odot$, for Z=0 and Z between $10^{-4}$ and Z$_\odot$. Si, O, Ne and 
C explosive burning occur when the shock wave crosses the corresponding zones 
in the pre-supernova (see above for the description of its structure).  
A brief description of the results concerning the synthesis of radioactive 
isotopes follows. 

\nil\ and \nim\ are produced when either oxygen or silicon-rich layers with 
low neutron excess (Y$_{\rm e} \gap 0.498$) are 
heated to temperatures above 4x$10^9$ K (explosive O- and Si-burning). 
They are produced whether the 
material ejected is alpha-rich or not, although \nim\ synthesis is favored in 
alpha-rich freeze-out; this happens when material, initially in nuclear 
statistical equilibrium (NSE) at relatively low density, is cooled so 
rapidly that 
the free alpha particles do not have time to merge via the 3$\alpha$ reaction 
and, therefore, matter cools down in the presence of a large concentration of 
$\alpha$-particles, which modify the final composition (with respect to the 
normal freeze-out). \tit\ is also produced during $\alpha$-rich freeze-out 
from NSE in the hottest and deepest layers ejected during the explosion. 
Therefore, the yields of these radioactive isotopes are very sensitive to the 
mass-cut location (Woosley and Hoffman 1991, Hoffman et al. 1995, 
Woosley and Weaver 1995, Timmes et al. 1996). For example, stars with masses 
larger than 30 M$_\odot$ don't eject any \nil\ (nor \nim\ and \tit) if 
the kinetic energy (at infinity) is around 1.2x$10^{51}$ erg. If this energy 
is enhanced, the mass-cut is lowered and some \nil\ (and \nim\ and \tit) are 
ejected. Ejected masses of \nil\ are around 0.1 M$_\odot$ and those 
of \tit\ between $\sim 10^{-5}$ and $10^{-4}$ M$_\odot$. Similar results are 
obtained by Thielemann et al. (1996), except for the larger amounts of \tit, 
probably because of the different way of simulating the explosion, which 
possibly injects larger entropy in the inner shells and favors a larger 
ejected mass and an enhanced $\alpha$-rich freeze-out (see, e.g., 
Aufderheide, Baron and Thielemann 1991 and Hoffman et al. 1999).    

\all\ is another important radioactive isotope which is produced in 
core collapse supernovae (and in other scenarios) through the 
$^{25}$Mg(p,$\gamma$) reaction. \all\ yields depend on 
pre-supernova evolution (H- and O-Ne burning shells) and on the explosion. 
Two factors enhance \all\ production during the explosion: explosive burning 
in O-Ne shells and $\nu$-spallation reactions on $^{20}$Ne, $^{16}$O, 
$^{23}$Na, $^{24}$Mg, which liberate protons that are captured by $^{25}$Mg. 
It is important to stress that another important long-lived radioisotope, 
\feh, is coproduced with \all\ in the same regions within SNII (this isotope 
is synthesized by neutron captures on $^{56,58}$Fe in the O-Ne burning 
shell and in the base of the He-burning shell, both pre-explosively and 
explosively). Therefore, 
these nuclei should have similar spatial distributions in the ejecta (Timmes 
et al. 1995). The \all/\feh\ ratio depends on the mass of the pre supernova: 
\all/\feh\ is larger than 1 for M larger than 25 M$_\odot$ and similar to 1 
for smaller masses. The typical yields of \all\ are $10^{-4}$ M$_\odot$ and 
those of \feh\ 4x$10^{-5}$ M$_\odot$.

The final yields depend on three aspects: 
presupernova evolution, explosion energy and details of the explosion 
mechanism (see Diehl and Timmes, 1998 and Thielemann 1999 for recent 
analyses). The main issues 
concerning presupernova evolution are those affecting general stellar 
evolution of low-mass stars, plus some specific ones relative to massive 
stars. For instance, the treatement of convection affects nucleosynthesis; in 
particular, convective burning in the O-shell of massive stars. Models with 
M=20 M$_\odot$ have deserved a particular attention for the theorists, since 
they are  crucial to understand the mixing of radioactive (and other) 
isotopes, like \nil, in supernova ejecta, which has been deduced from the 
observations of SN1987A (see section 2.1 above). 2D models of O-burning 
(Bazan and Arnett 1994) obtain significant mixing beyond the boundaries 
defined by 
mixing-length convection. What they obtain are perturbations in density 
in the oxygen shell that are sufficiently large to ``seed'' hydrodynamic
instabilities, which will mix the ``onion-skin'' composition of the 
presupernova (Bazan and Arnett 1998). This occurs in precisely the region 
in which \nil\ is explosively produced by oxygen burning behind the 
explosion shock. This result poses some problems to the models of explosive 
nucleosynthesis based on 1D presupernova evolution. Rotation can also have 
some effect (see, e.g., works by Meynet and Maeder 1997, Heger, Langer and 
Woosley 2000), as well as mass-loss during the presupernova evolution, in the 
final yields of radioactive elements. Concerning the energy of the explosion 
and the details of the explosion mechanism, one of the main problems is the 
location of the mass-cut, which determines how much mass falls back into the 
collapsing core (and therefore whether it will be a neutron star or a black 
hole) and how much mass is ejected and with wich composition (the profile 
of some isotopes is steep around the mass-cut location and, therefore, the 
yield is affected by it). Therefore, the mass-cut determines crucially the 
final yields of radioactive elements, specially for those produced in the 
inner regions of the supernova (\nil, \nim\ and \tit). As mentioned above, 
the explosion energy, and the corresponding entropy in the inner shells, 
crucially affect the degree of  $\alpha$-rich freeze-out and, therefore, 
the yields of the Fe-group nuclei and of \tit.  

\section{Classical novae}
Classical novae explosions are the most common explosions in the Galaxy. The 
cause of the explosion is a thermonuclear runaway (TNR) on top of a white 
dwarf, ensuing the degenerate burning of the accreted hydrogen (Starrfield 
1989, Hernanz \& Jos\'e 2000). 
The synthesis of radioactive isotopes in classical novae is important for two 
reasons. First, some of the isotopes produced are crucial for the 
explosion mechanism itself (i.e., $^{14}$O, $^{15}$O, $^{17}$F with lifetimes 
102, 176 and 93s, respectively). The reason is that these isotopes are 
transported by 
convection to the outer layers of the envelope, during the runaway, where 
they subsequently decay ($\tau_{\rm conv} < \tau_{\rm decay}$) and cause 
the expansion of the envelope and the increase in visual luminosity. 
Second, the 
decay of the unstable nuclei originates gamma-ray emission, because of 
either direct emission of gamma-ray photons or positrons (for 
$\beta^{+}$-unstable nuclei), which annihilate with electrons. The photons 
emitted 
(511 keV, positronium continuum, 478 and 1275 keV, see table 1) experience 
Comptonization in the nova expanding envelope. Therefore, the emission from 
novae consists of lines plus a continuum (see G\'omez-Gomar et al. 1998b, 
Hernanz et al. 1999, 
Hernanz et al. these proceedings and references therein). The potential role 
of classical novae as gamma-ray emitters had been alredy pointed out many 
years ago (Clayton and Hoyle 1974, Clayton 1981, Leising and Clayton 1987).

The first available hydrodynamic models of nova explosions (Starrfield et al. 
1978 and Prialnik et al. 1978) realized that there was a need of an initial 
enrichment in CNO isotopes both to power the explosion and to explain 
some observed abundances. Two and three dimensional simulations of the 
thermonuclear runaway of a CO white dwarf, valid when the accreted envelope 
has been already built up, are the only available up to now 
(Glasner et al., 1997, Kercek et al. 1998, 1999). They predict that 
enrichment proceeds too slowly if the accreted gas has nearly solar 
CNO abundances at the onset of the thermonuclear runaway, 
and conclude that fast nova outbursts require huge enrichments of C and O. 
The mechanism
which leads to such enhancements must operate prior to the outburst and has 
not been modeled up to now.
Therefore, it is known that some mixing with core material (either CO or ONe) 
during the accretion phase 
prior to the runaway should occur, but this process has not been modeled yet 
in a self-consistent way. Another approach to the problem of initial 
enrichment comes from the multicycle 1D models (of CO novae only, up to now), 
from Prialnik and Kovetz (Prialnik and Kovetz 1995, Kovetz and Prialnik, 
1997); diffusion is responsible for the enrichment, which becomes larger 
after a number of flashes. However, the large metallicities and neon 
abundances 
observed in some novae are not well reproduced. Another approach is based on 
1D models with an initial (parametrized) enrichment, such that the general 
properties of observed novae (mainly abundances) are well modeled (see for 
instance Starrfield et al. 1998, Jos\'e and Hernanz 1998).

Classical novae synthesize many radioactive isotopes, which vary depending on 
the nova type (which in turn depends on the type -CO or ONe- of the 
underlying white dwarf). CO novae produce mainly \bee, whereas ONe produce 
\naa\ and \all. Other radioactivities with shorter lifetimes, such as 
$^{13}$N and $^{18}$F ($\tau=862$s and 158min, respectively) are produced in  
similar amounts in CO and ONe novae (see Jos\'e and Hernanz, 1998, Jos\'e, 
Coc and Hernanz, 1999, Hernanz et al. 1999 and Hernanz et al., these 
proceedings for details). Typically, $10^{-7}-10^{-8}$M$_\odot$ of $^{13}$N, 
$10^{-9}$M$_\odot$ of $^{18}$F are ejected in both CO and ONe explosions. 
CO novae also eject $10^{-10}$M$_\odot$ of \bee, and ONe novae 
$10^{-9}$M$_\odot$ of \naa\ and $10^{-8}$M$_\odot$ of \all. 
The reason of the different explosive nucleosynthesis results in CO and ONe 
nova types is that some seed nuclei (such as $^{20}$Ne, \neo, $^{24,25}$Mg) 
are necessary to synthesize \naa\ and \all. That's because temperatures 
attained at the peak of the nova outburst are not high enough to break the 
CNO cycle towards NeNa-MgAl cycles. 

The two short-lived isotopes $^{13}$N and $^{18}$F are crucial for the prompt 
gamma-ray emission of novae, which is the most intense emission ($10^{-3}$ 
phot~cm$^{-2}$~s$^{-1}$), but of very short duration (a few days) and 
appearing 
before optical detection. The medium-lived isotopes (\bee\ and \naa) produce 
fluxes of around $10^{-6}$ and $10^{-5}$ phot~cm$^{-2}$~s$^{-1}$, for 
distances of 1 kpc. The prospects for detectability with the SPI instrument 
onboard INTEGRAL are analyzed in Hernanz et al. (these proceedings, and 
references therein). It is important to remind that, in addition to the 
\bee\ and \naa\ emission from individual novae, the cumulative emission from 
all the galactic novae can give important information about the distribution 
of the sources, specially if there is only one dominant source for that 
particular isotope. For \naa, novae are the main individual contributors. 
Therefore, the detection of galactic \naa\ emission, and the corresponding 
1275 keV emission map (see Jean et al., 1999, 2000), would be a very valuable 
tool 
to study the distribution of novae in the Galaxy, which is very poorly 
known from optical-UV and IR observations because of interstellar extinction.
Finally, \all\ is produced in ONe novae in such an amount 
that makes it quite improbable that novae contribute largely to the 
\all\ content of the Galaxy, as observed through its emission at 1809 keV.

\section*{Acknowledgments}
Research partially supported by the CICYT-P.N.I.E.
(ESP98-1348), by the DGES (PB98-1183-C03-02 and PB98-1183-C03-03) and 
by the AIHF1999-0140


\begin{table*}
\begin{center}
\caption{Radioactive isotopes synthesized in explosive events}
\vspace{0.3cm}
\renewcommand{\arraystretch}{1.5} 
\begin{tabular}{cccc}
\hline
Isotope     & Decay chain 
            & Lifetime                  & Line energy (keV) \\
\hline
\nil\       & \nil \  $\rightarrow$ \ \col                
            & 8.8d                      & 158, 812, 750, 480 \\
\col\       & \col \  $\rightarrow$ \ \fel                
            & 111d                      & 847, 1238 \\
\nim\       & \nim \  $\rightarrow$ \ \com \  $\rightarrow$ \ \fem  
            & (52h) 390d                & 122, 136 \\
\tit\       & \tit \  $\rightarrow$ \ \sca \  $\rightarrow$ \ \caa  
            & 89yr (5.4h)               & 78, 68, 1157 \\
\all\       & \all \  $\rightarrow$ \ \mgg                
            & 1.0x$10^6$yr              & 1809 \\
\feh\       & \feh \  $\rightarrow$ \ \coh \  $\rightarrow$ \ \nih  
            & 2.0x$10^6$yr (7.6yr)      & 1173, 1332 \\
\bee\       & \bee \  $\rightarrow$ \ \lit                
            & 77d                       & 478 \\
\naa\       & \naa \  $\rightarrow$ \ \neo                
            & 3.8yr                     & 1275 \\
\hline
\end{tabular}
\end{center}
\end{table*}

\end{document}